\documentclass[aip,reprint]{revtex4-1}
\usepackage[T1]{fontenc}
\usepackage{graphicx}
\usepackage{subfig}
\usepackage{xcolor}
\usepackage{upgreek}
\usepackage{tabularx}
\usepackage{ulem}
\begin{document}
\title{Molecular rotors to probe the local viscosity of a polymer glass} 
\author{Elham Mirzahossein}
\email{e.mirzahossein@uva.nl}
\affiliation{Van der Waals-Zeeman Institute, Institute of Physics, University of Amsterdam, 1098XH Amsterdam, The Netherlands}
\author{Marion Grzelka}
\affiliation{Van der Waals-Zeeman Institute, Institute of Physics, University of Amsterdam, 1098XH Amsterdam, The Netherlands}
\author{Zhongcheng Pan}
\affiliation{Van der Waals-Zeeman Institute, Institute of Physics, University of Amsterdam, 1098XH Amsterdam, The Netherlands}
\author{Beg\"{u}m Demirkurt}
\affiliation{Van 't Hoff Institute for Molecular Sciences, University of Amsterdam, 1098XH Amsterdam, The Netherlands}
\author{Mehdi Habibi}
\affiliation{Physics and Physical Chemistry of Foods, Wageningen University \& Research, 6708 WG Wageningen, Wageningen, The Netherlands}
\author{Albert M. Brouwer}
\affiliation{Van 't Hoff Institute for Molecular Sciences, University of Amsterdam, 1098XH Amsterdam, The Netherlands}
\author{Daniel Bonn}
\affiliation{Van der Waals-Zeeman Institute, Institute of Physics, University of Amsterdam, 1098XH Amsterdam, The Netherlands}
\date{\today}
\begin{abstract}
We investigate the local viscosity of a polymer glass around its glass transition temperature using environment-sensitive fluorescent molecular rotors embedded in the polymer matrix. The rotors' fluorescence depends on the local viscosity, and measuring the fluorescence intensity and lifetime of the probe therefore allows to measure the local free volume in the polymer glass when going through the glass transition. This also allows us to study the local viscosity and free volume when the polymer \textcolor{black}{film} is put under an external stress.We find that the \textcolor{black}{film} does not flow homogeneously, but undergoes shear banding that is visible as a spatially varying free volume and viscosity.\\

Published in J. Chem. Phys.\\
\doi{10.1063/5.0087572}
\end{abstract}
\maketitle 
\section{Introduction}

Polymer glasses are ubiquitous in our everyday life. They are typically made by heating a polymer  above its glass transition temperature, so that it can be extruded, cast or molded. When subsequently cooled below the glass transition temperature, a hard and sometimes even brittle solid results. For instance,
\textcolor{black}{hard plastics like poly(methyl methacrylate) (plastic and bulletproof windows) and polystyrene (car instrument panels) , with $T_\mathrm{g}$ values of about 100$^\circ$C or polyvinyl chloride (plumbing pipes) with $T_\mathrm{g}$ values of about 80$^\circ$C are well below their glass transition temperatures $T_\mathrm{g}$ at room temperature \cite{kuo_thermal_2003,alford_specific_1955,rudin_effects_1975}}. The transition from a liquid-like state to a mechanically solid state is known as the glass transition that is generically characterized by 
the change in mechanical properties around the glass transition~\cite{pan_conformational_2008,cangialosi_dynamics_2014,vidal_russell_direct_2000}. For instance, the IUPAC definition of the glass transition temperature is the temperature at which the viscosity of the glass is $10^{12}\,$Pa s. While hard plastics are used \textcolor{black}{well below ($\geq 50\,$ K)} their glass transition temperatures, elastomers are usually employed above their $T_\mathrm{g}$, in their rubbery state. This is what makes rubbers soft and flexible; for most rubbers interchain crosslinking prevents large-scale rearrangements of the polymers, so that even above their $T_\mathrm{g}$ they do not flow. 

However, due to its complexity, the microstructure of polymer glass around $T_\mathrm{g}$ is not well understood.  The presence of dynamic heterogeneity is believed to play a key role in the relaxation and mechanical properties of glasses~\cite{balbuena_structural_2021,a_riggleman_heterogeneous_2010,yoshimoto_mechanical_2004}. Nevertheless, local measurements of dynamic heterogeneity are \textcolor{black}{limited}, because it is difficult to measure at small scales, and because relaxation is extremely slow below $T_\mathrm{g}$. Generically, conformational changes of the polymer segments, typically consisting of 10–20 monomers are believed to become extremely slow below the glass transition temperature: the segmental relaxation is almost completely halted, and this would be the first step in locally reorganizing a polymer glass. However, everyday experience shows that if sufficient stress is applied to a glassy polymer, such as during extrusion, molding or casting of polymeric materials, it can deform significantly, involving consequent movement of polymer chains. Again, the behavior of the (local) microstructure under high stress remains poorly understood~\cite{siekierzycka_polymer_2010,ellison_confinement_2002,willets_novel_2003}. In their pioneering work, Ediger and colleagues measured changes in molecular mobility during the deformation of a glassy polymer, and have shown that under stress, segmental mobility may increase by up to a factor of 1000.~\textcolor{black}{\cite{lee_dye_2008,lee_molecular_2009,lee_direct_2009,lee_interaction_2010,hebert_reversing_2017,bending_comparison_2016,lee_deformation-induced_2009,riggleman_free_2007,hall_translation-rotation_1997}.}  \textcolor{black}{Using photobleaching techniques, they have also shown that in the early stage of deformation, polymer glass presents strong dynamic heterogeneity and becomes more homogeneous after the onset of the flow~\cite{a_riggleman_heterogeneous_2010}. Moreover, they observed the dynamics of full recovery of the polymer glass after removing the extensional stress~\cite{lee_mechanical_2010}.}

An additional complication is that viscoelastic materials such as yield stress fluids, micellar solutions~\cite{salmon_velocity_2003,radulescu_time_2003} and glassy materials in general ~\cite{varnik_study_2004} exhibit shear banding, i.e., they do not flow homogeneously across the sample thickness. This phenomenon also occurs for polymer glasses \textcolor{black}{and polymer melts} under stress: the deformation in polymer \textcolor{black}{film} is not uniformly distributed through the sample and tends to accumulate close to a wall in a shear band~\cite{tapadia_banding_2006,cao_shear_2012,wisitsorasak_dynamical_2017,ravindranath_banding_2008} 
Understanding the formation of both dynamic and mechanical heterogeneity in polymer glasses around their glass transition is one of the central questions in glass science. This calls for spatially resolved local measurements of the polymer dynamics under stress, which is not easy to achieve.

\textcolor{black}{In this paper} we attempt such a measurement, and show the promise that local viscosity probes can have for the study of glassy dynamics. We look at the local changes in polymer glasses at the microscopic scale using so-called molecular rotors. These are fluorescent dyes that are sensitive to their molecular environment with 
a high sensitivity and both high spatial and temporal resolutions. Molecular rotors have a viscosity-dependent non-radiative excited state decay pathway (rate $k_\mathrm{nr}$) that can out-compete fluorescence (rate $k_\mathrm{f}$) in low viscosity media, leading to low fluorescence quantum yield $\Phi_\mathrm{f}$ and short decay time $\tau_\mathrm{f}$. At higher microviscosities, $k_\mathrm{nr}$ is smaller, and both $\Phi_\mathrm{f}$ and $\tau_\mathrm{f}$ increase.
Therefore, the fluorescence quantum yield of molecular rotors strongly depends on the nature of the direct environment of the dye, especially the microviscosity. \textcolor{black}{Molecular rotors have already been extensively used to probe polymer matrices; using spectrofluorimetry techniques, Loutfy \textit{et al.} studied the effect of polymer chain tacticity~\cite{loutfy_effect_1983}, side groups~\cite{law_fluorescence_1983} or the polymerization process~\cite{loutfy_high-conversion_1981}. Numerous works used molecular rotors fluorescent properties to study the effect of confinement on the glass transition temperature~\cite{ellison_confinement_2002,priestley_effects_2007,kim_effect_2008,kim_confinement_2009}. Moreover, Loufty \textit{et al.} pioneer studies link the free volume to the fluorescence properties of molecular rotors~\cite{loutfy_effect_1982,law_spectroscopy_1981}, with a relationship that can become complex below the glass transition temperature~\cite{hooker_coupling_1995}. Such dyes have proven to be useful to probe dynamics in thin polymer films, to study physical aging~\cite{royal_monitoring_1990,royal_molecular-scale_1992,royal_physical_1993, ellison_confinement_2002}. Note that all these studies probes the polymer glass in bulk without any information of the homogeneity of samples at the micrometric scales.}

\begin{figure}%
    \centering
    \includegraphics[width=\columnwidth]{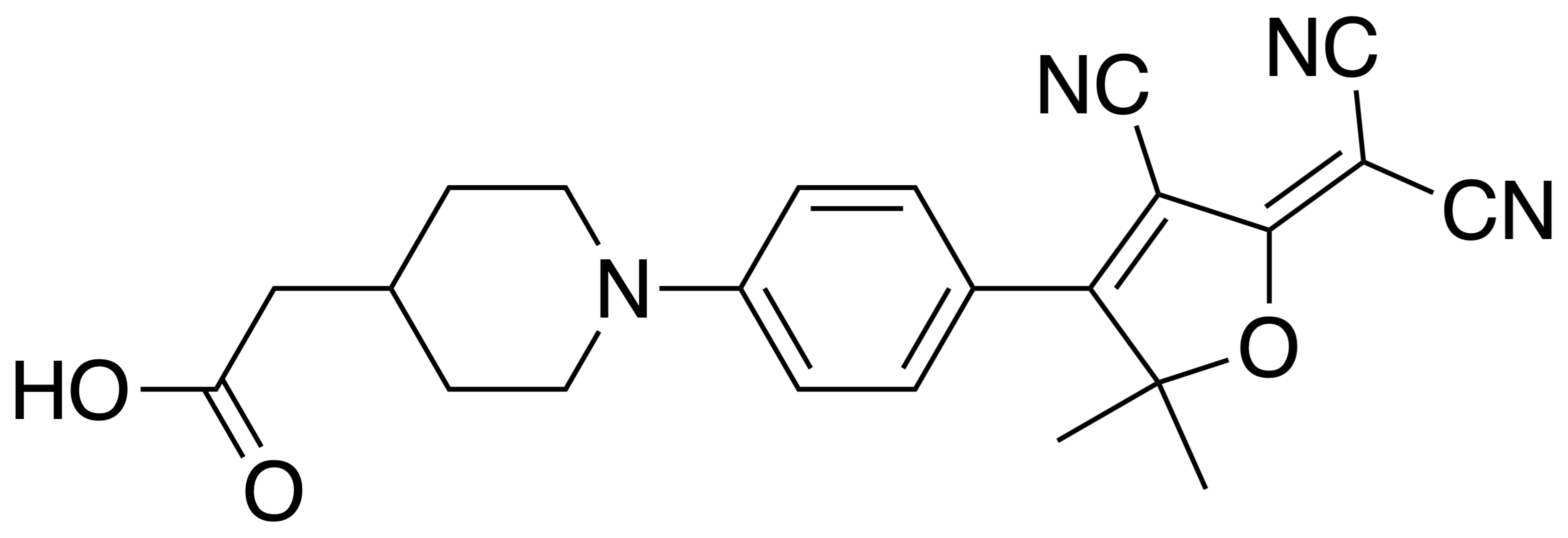} %
    \caption{Chemical structure of dicyanodihydrofuran (DCDHF) molecular rotors.}%
    \label{fig:DCDHF}%
\end{figure}

\textcolor{black}{Here, we use the dicyanodihydrofuran chromophore (DCDHF, see Fig.\,\ref{fig:DCDHF}) dye to measure the  \textit{local} viscosity in a polymer glass, which can be related to free volume changes. This is done by 3D confocal fluorescence microscopy, so that we can resolve spatial changes in fluorescence intensity or lifetime in the $x$, $y$ and $z$ directions.} The dicyanodihydrofuran chromophore (DCDHF, see Fig.\,\ref{fig:DCDHF}) that we use here is one example of such molecular rotors: it has been used as a probe in many studies such as fluorescence sensors for contact mechanics measurements~\cite{suhina_photophysics_2021,weber_molecular_2018} or single-molecule imaging in cells~\cite{lord_dcdhf_2009}. The molecular rotors' sensitivity to its local environment has already been used to probe the local viscosity of viscoelastic media~\cite{jee_internal_2010,jee_determination_2013}. 
We probe a polyvinyl acetate (PVAc) polymer glass around its glass transition temperature. We observe that, as expected, the free volume of the PVAc polymer glass increases with increasing temperature around glass transition temperature, and find that the fluorescence is continuously changing across the glass transition. Moreover, we performed a simple squeezing experiment \textcolor{black}{on PVAc polymer film at temperatures $T_\mathrm{g} +10\,$K and  $T_\mathrm{g}+15\,$K.} We measured a decrease in fluorescence intensity characteristic of a clear shear banding effect in the $z$ (gravity) direction, with roughly one half of the sample having a very different local viscosity than the other half. Remarkably, even under stress, the fluorescence is very homogeneous at the microscopic scale in the ($x,y$) plane, showing no large effects of dynamic heterogeneity.

\section{Material and Methods}
\subsection{Materials}
PVAc glasses are prepared for two different molar weights ($M_\mathrm{w}\sim100\,$kg/mol from Sigma-Aldrich and $M_\mathrm{w}\sim500\,$kg/mol from Scientific Polymer Products). We briefly detail the protocol to embed the molecular rotor in the polymer glasses. DCDHF fluorescent probes (available from previous work~\cite{suhina_photophysics_2021}) are added to a solution of PVAc in tetrahydrofuran (THF from Sigma-Aldrich) of concentration $20 $ wt\%. 
The PVAc glass film is prepared by drop-casting this polymer solution on a plasma-cleaned glass slide. Total evaporation of THF is ensured by vacuum for 2 days and drying in an oven at $70 ^{\circ}$C for another 3 days. The total concentration of DCDHF probes in PVAc glass is chosen to be $ 10^{-8}\,$mol/L to avoid any dye-dye interaction in the sample. The film thickness $h_\mathrm{film}$ and diameter $d_\mathrm{film}$ vary respectively between 60 and 250 $\upmu$m and 5 and 7 mm. For most of the experiments presented here, having thicker samples simplifies the measurements: we take two glass slides with a drop cast on it, and melt these together to get a glassy polymer bridge between two glass slides. \textcolor{black}{PVAc samples are kept in a dry environment and are used at ambient relative humidity.} The temperature of the sample is controlled in the range $ 20^{\circ}$C to $50^{\circ}$C thanks to a home-made Peltier plate placed above the sample.

\subsection{Methods}
\subsubsection{Confocal microscopy}
\begin{figure}%
    \centering
    \includegraphics[width=\columnwidth]{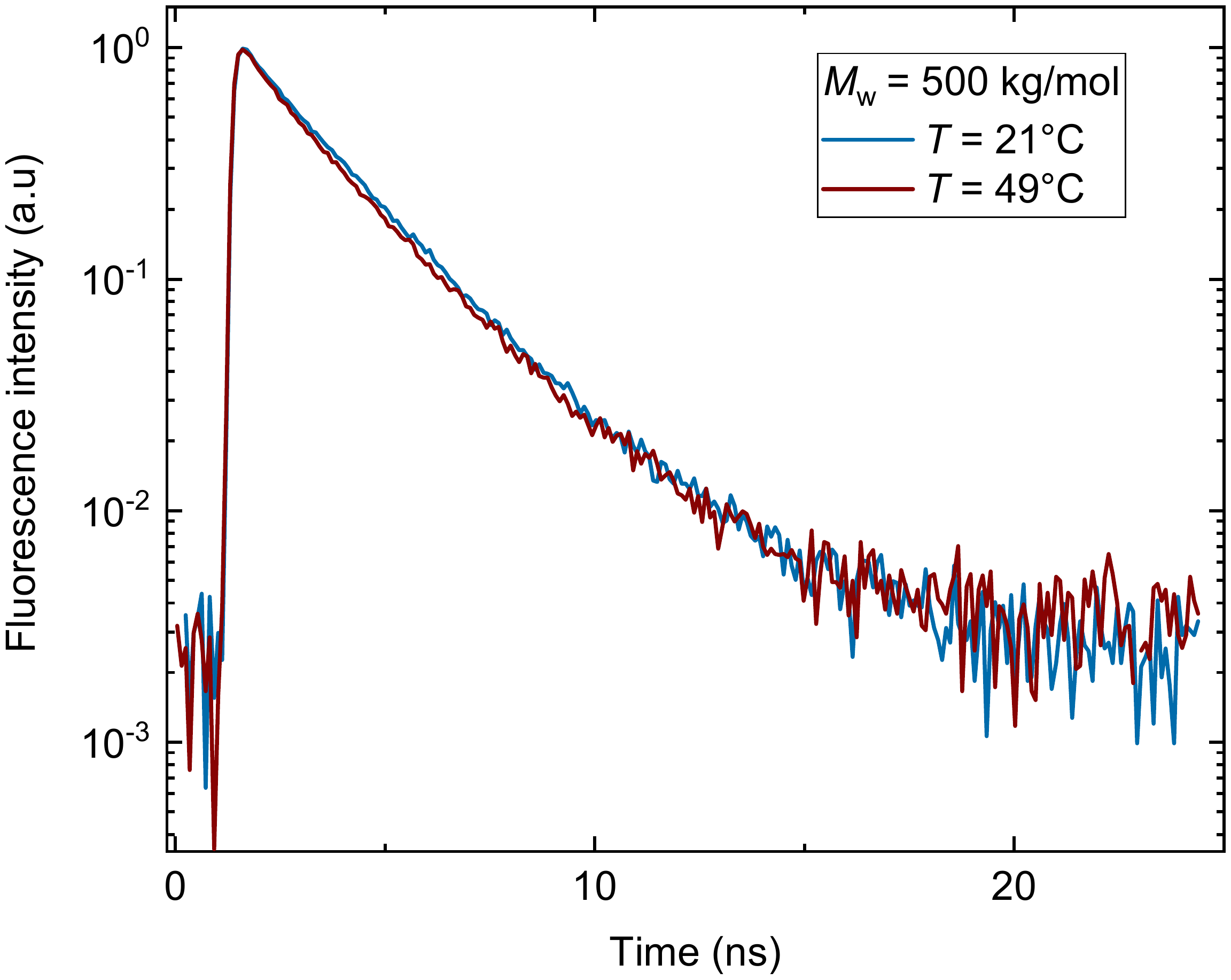} %
    \caption{Time-resolved fluorescence decays of DCDHF in PVAc polymer glass ($M_\mathrm{w}\sim500\,$kg/mol) at $T = 21 ^\circ$C (blue line) and $T = 49 ^\circ$C (red line).}%
    \label{fig:Decay_curve}%
\end{figure}

The Fluorescence Lifetime Imaging (FLIM) and intensity of the molecular probes are measured using an inverted confocal microscope (Leica TCS SP8) equipped with a hybrid detector. We used a 20X dry objective with a numerical aperture of 0.75. 
A Falcon photon counting module, directly coupled to the microscope is used for lifetime measurements. For lifetime measurements, the excitation wavelength of the pulsed laser (repetition rate of 40 MHz) is 470 nm and the emission wavelength range is between 500 and 700 nm. On the other hand, for the fluorescence intensity measurements, the excitation wavelength is 488 nm and the emission wavelength range is chosen between 510 and 700 nm. The resolution in the $z$ (gravity) direction is set by the optical sectioning, roughly estimated to be $2\,\upmu$m with the used wavelength, pinhole and objective. The laser power is set at the lowest value to avoid photobleaching but still have enough fluorescence signal: for lifetime measurements, we set the laser power at $\sim21\,\upmu$W and for intensity measurements respectively at $\sim40\,\upmu$W for $M_\mathrm{w}\sim500\,$kg/mol and $\sim 110\,\upmu$W for $M_\mathrm{w}\sim100\,$kg/mol. All the measurements have been done from high temperature to low temperature, to avoid the possible effect of photobleaching to change the observed behavior. Figure \ref{fig:Decay_curve} presents a typical Time Correlated Single Photon Counting (TCSPC) histogram for DCDHF embedded in a PVAc glass: TCSPC is bi-exponential for all the measurements. This reflects an heterogeneous environment at the scale probed here, typically limited by the lateral resolution, roughly estimated to be $400\,$nm. In the following, all the reported lifetime values are the amplitude average lifetime $\langle \tau \rangle$ defined as:
\begin{equation}
\langle \tau \rangle = \frac{A_1  \tau_1 +A_2  \tau_2}{A_1 +A_2}
\label{eq:A_LF}
\end{equation}
with $A_i$ and $\tau_i$ are the amplitude and lifetime of the i-th decay component. If the data presented in Fig.\,\ref{fig:Decay_curve} seem similar for $T=21^\circ$C and $T=49^\circ$C, the bi-exponential fits allow to distinguish them, with respectively $A_1=7.9$, $A_2=7.6$, $\tau_1=1.54\,$ns and $\tau_2=2.60\,$ns for $T=21^\circ$C and $A_1=4.8$, $A_2=11.4$, $\tau_1=0.90\,$ns and $\tau_2=2.33\,$ns for $T=49^\circ$C: this leads to slightly different calculated amplitude average lifetime $\langle \tau \rangle =2.06\,$ns for $T=21^\circ$C and $\langle \tau \rangle =1.91\,$ns for $T=49^\circ$C.

\subsubsection{Rheological measurement}

The rheological properties of PVAc glasses were measured by oscillatory rheology at different temperatures using an Anton-Paar MCR 302 rheometer in a parallel plate geometry (9 mm diameter). The polymer film is 'glued' to the plate by melting the film using a hot air gun ($\sim200^{\circ}$C). The film thickness and diameter are taken into account in interpreting the rheology data.

\subsubsection{Squeezing experiment}
Squeezing experiments were performed by applying a normal force $F=10\,$N to the polymer films using an Anton Paar MCR 301 rheometer. \textcolor{black}{Such a normal force corresponds to a stress value $\sigma = \frac{F}{A_0}\sim 0.8\,$MPa applied to a polymer film, where $A_0$ is the cross section area of the center of the sample.} The gap size variation is measured during the squeezing at a fixed temperature ($T=50$ and $45^\circ$C). Shortly after the squeezing (typically 10 min), the ($x$,$y$,$z$) changes in fluorescence intensity are measured on the inverted confocal microscope at the same temperature. Note that the time between the end of the squeezing and the fluorescence measurements is too short to observe relaxations in the polymer glass.


\subsubsection{\textcolor{black}{Differential scanning calorimetry}}

\textcolor{black}{Differential scanning calorimetry (DSC) measurements were performed on a TA Instruments DSC Q100. PVAc ($M_\mathrm{w}\sim500\,$kg/mol) is heated from $-10^\circ$C to $80^\circ$C at a heating rate of $10^\circ$C/min followed by an isothermal step for 1 min. A cooling cycle to $-10^\circ$C at a rate of $10^\circ$C/min was performed prior to a second heating run to $80^\circ$C at the same heating rate. The glass transition temperature determined by DSC was defined as the temperature of the midpoint of a heat capacity change on the second heating run. The Universal Analysis 2000 software was used for data acquisition.}


\section{Experimental results and discussion}

\subsection{Glass transition induced by temperature}
\begin{figure}[t]%
    \centering
    \includegraphics[width=\columnwidth]{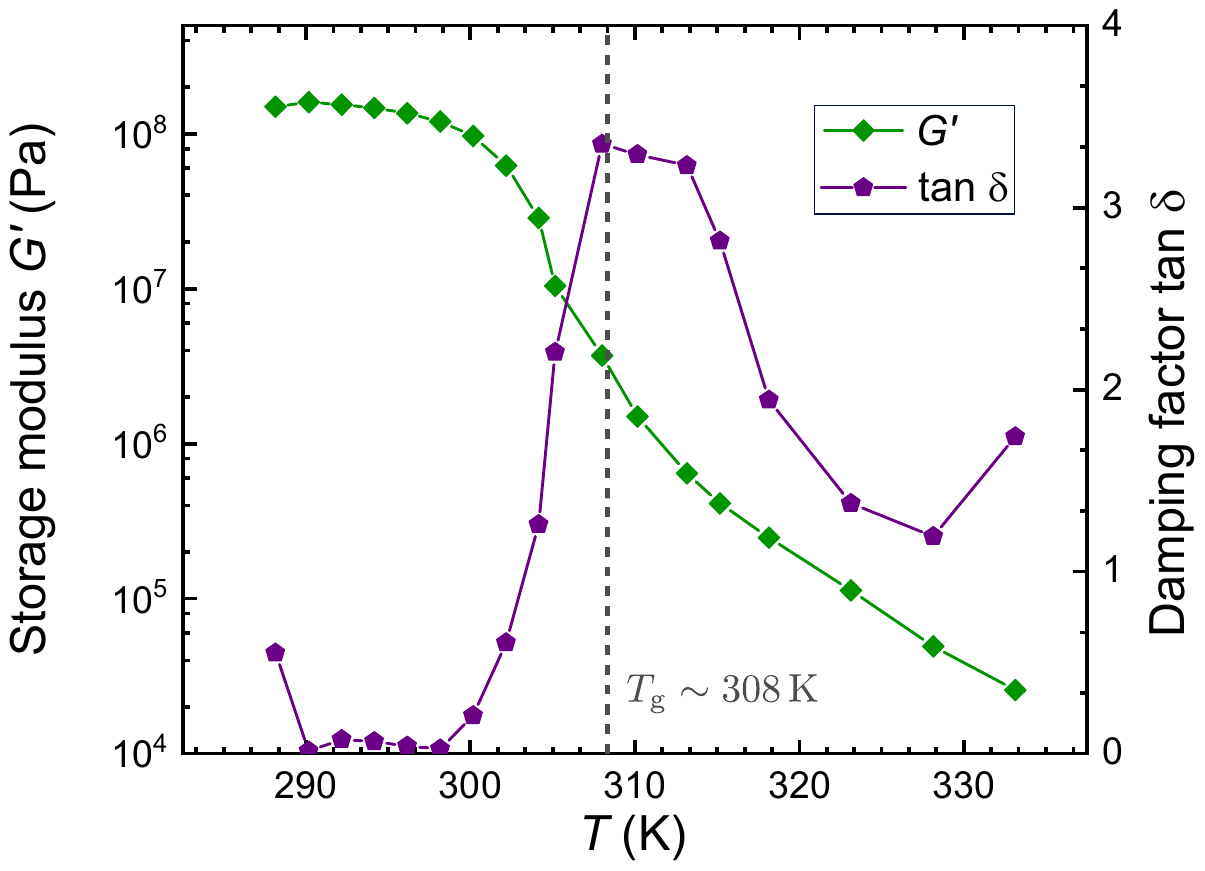} %
    \caption{Storage modulus $G^\prime$ and damping factor $\tan\delta = \frac{G^{\prime \prime}}{G^\prime}$ as a function of temperature $T$ for a PVAc polymer glass ($M_\mathrm{w} \sim 500\,$kg/mol, $h_\mathrm{film}=250\,\upmu$m and $d_\mathrm{film}=6.5\,$mm). $T_\mathrm{g}\sim308\,$K is determined by local maximum value of $\tan\delta$}
    \label{fig:RheologyvsT}
\end{figure}

The storage modulus $G^\prime$ and loss modulus $G^{\prime\prime}$ of a PVAc film are measured by applying small-amplitude oscillatory shear with the strain amplitude $\gamma_0$ set at $0.1\,\%$ and the frequency set at $0.01\,$Hz at different temperatures around glass transition temperature. The storage modulus $G^\prime$ exhibits a plateau at low temperatures and then abruptly decreases (rubbery state) when going through the glass transition temperature (Fig.\,\ref{fig:RheologyvsT}). The value of the glass transition temperature $T_\mathrm{g} \sim 308\,$K (for $M_\mathrm{w} \sim 500\,$kg/mol) is determined by the local maximum value of damping factor $\tan\delta = \frac{G^{\prime \prime}}{G^\prime}$. \textcolor{black}{The glass transition temperature measured by the local peak in $\tan\delta$ is slightly overestimated compared to the one measured by DSC, $T_\mathrm{g,\,DSC} = 305.8\,$K. In the following, both determinations of the glass transition temperature will be discussed.}

\begin{figure}%
    \centering
    \includegraphics[width=\columnwidth]{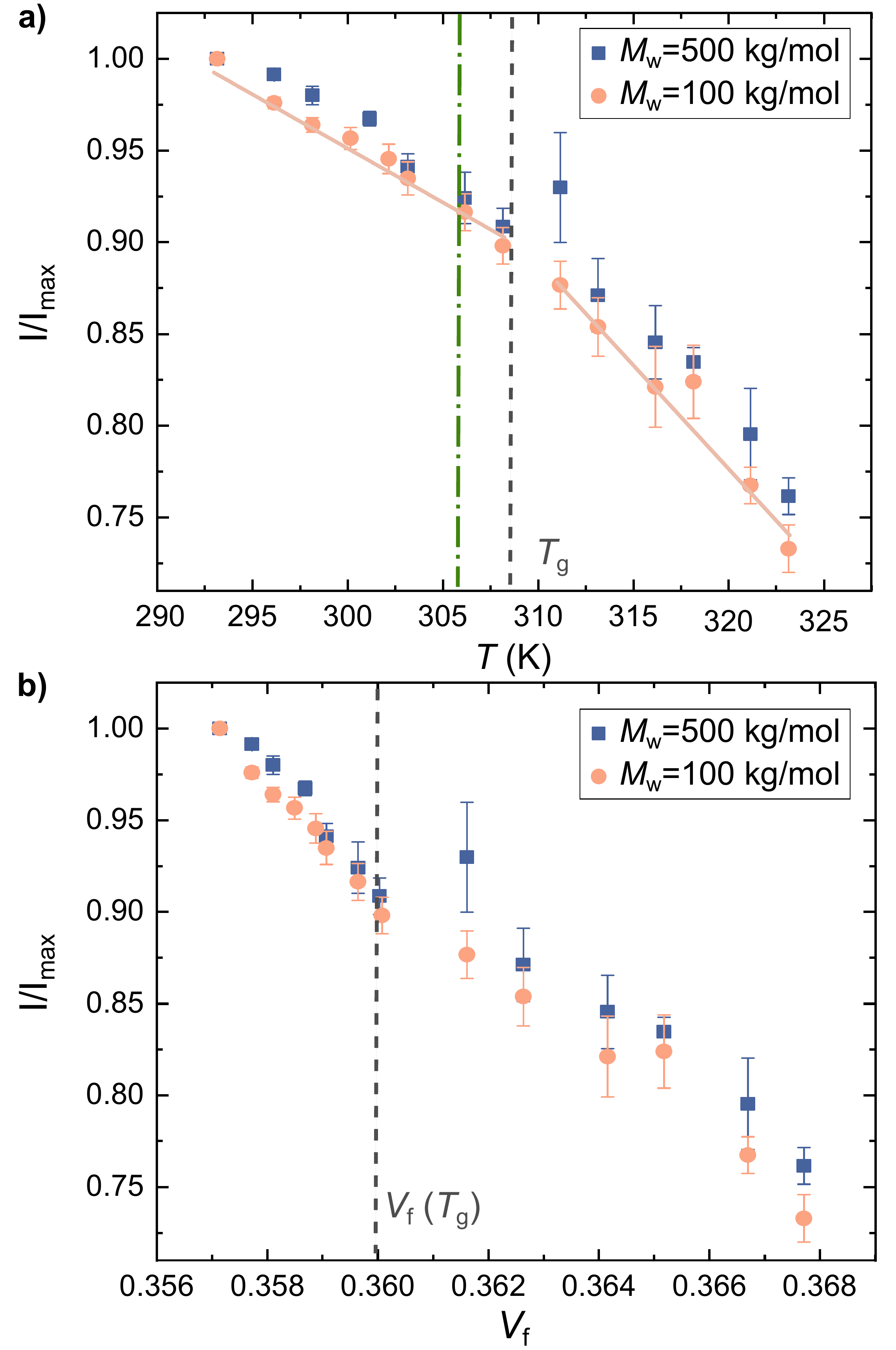}
    \caption{Normalized intensity of DCDHF in PVAc polymer glass with two different molecular weight as a function of (a) temperature $T$ and (b) free volume $V_\mathrm{f}$. \textcolor{black}{The green and black dashed line is glass transition temperature measured by differential scanning calorimetry and rheological measurements, respectively. The solid lines are guides for the eye.}}%
    \label{fig:PVAc_Intensity}%
\end{figure}

We subsequently measure the fluorescence response of DCDHF probes inside the PVAc polymer glass as a function of temperature. We perform all the experiments on the two different molecular weights of PVAc polymer glass that have almost the same glass transition temperature\cite{saito_temperature_1963}.
$I_\mathrm{max}$ is defined as the maximum intensity, corresponding to the intensity measured at the lowest temperature. The error bars in all the measurements are standard error resulting from three different positions in the ($x,y$) plane. As shown in Fig.\,\ref{fig:PVAc_Intensity}a, we observe that the normalized intensity $(I/I_\mathrm{max})$ decreases with increasing temperature, for both studied molecular weights. This decrease of intensity finds it origin in the thermal expansion of the polymer glass, that leads both to the local increases of the free volume around the probe on a nanoscopic scale, and a concomitant decrease of the local viscosity. In other words, by changing the temperature (around the glass transition temperature), the local free volume increases, which causes enhancing the molecular mobility of the probes: the non-radiative decay is more efficient and thus DCDHF probes fluoresce less. We find that for the DCDHF probes, the normalized intensity significantly drops by about 25\,\% in a rather narrow temperature range around the glass transition temperature. \textcolor{black}{Such a decrease is similar to the one reported by Torkelson \textit{et al.}\cite{ellison_confinement_2002,kim_confinement_2009} for pyrene embedded in PS, PiBMA, and P2VP films (a $\sim40\,\%$ decrease in intensity over 80K). 
Moreover, fluorescence intensity decreases with the temperature faster above the glass transition temperature (see the solid lines to guide the eye in Fig.\,\ref{fig:PVAc_Intensity}a)).} This indicates that the change of intensity as a function of temperature is more important above $T_\mathrm{g}$ rather than below probably due to higher mobility of the dye in the polymer glass. \textcolor{black}{Interestingly, the change of slope in the decrease appears closer to the glass transition temperature measured by rheology than to the one of DSC. We therefore use the rheology determination $T_\mathrm{g}\sim308\,$K in the rest of the article.}


The temperature-induced polymer glass transition is due to the shrinkage of the whole material at low $T$ given by the thermal expansion coefficien $\alpha_T$, and the free volume fraction $V_\mathrm{f}$ which is described by the theory of Zaccone \textit{et al.}~\cite{zaccone_disorder-assisted_2013}:
\begin{equation}
    V_\mathrm{f} = 1- \phi = 1- \phi_\mathrm{c} e^{\alpha_T (T_\mathrm{g} - T)}
\end{equation}
where $\phi$ is the monomer packing fraction with a critical packing fraction $\phi_\mathrm{c}$ at $T_\mathrm{g}$. If we treat the monomers like effective hard spheres, the packing fraction would have the value of $\phi_\mathrm{c} = 0.64$. For PVAc, the thermal expansion coefficient $\alpha_T$ is equal to $ 3 \times 10^{-4} \mathrm{K}^{-1}$ for $T< T_\mathrm{g}$ and $ 8 \times 10^{-4}\mathrm{K}^{-1}$ for $T> T_\mathrm{g}$. \cite{bogoslovov_effect_2008}. 

Using of this theory we can directly connect the temperature to free volume in the polymer glass. 
If we plot the normalized intensity this time as a function of free volume (Fig.\,\ref{fig:PVAc_Intensity}b), we indeed observe the decrease of intensity as a function of free volume.

\begin{figure}[t]%
    \includegraphics[width=\columnwidth]{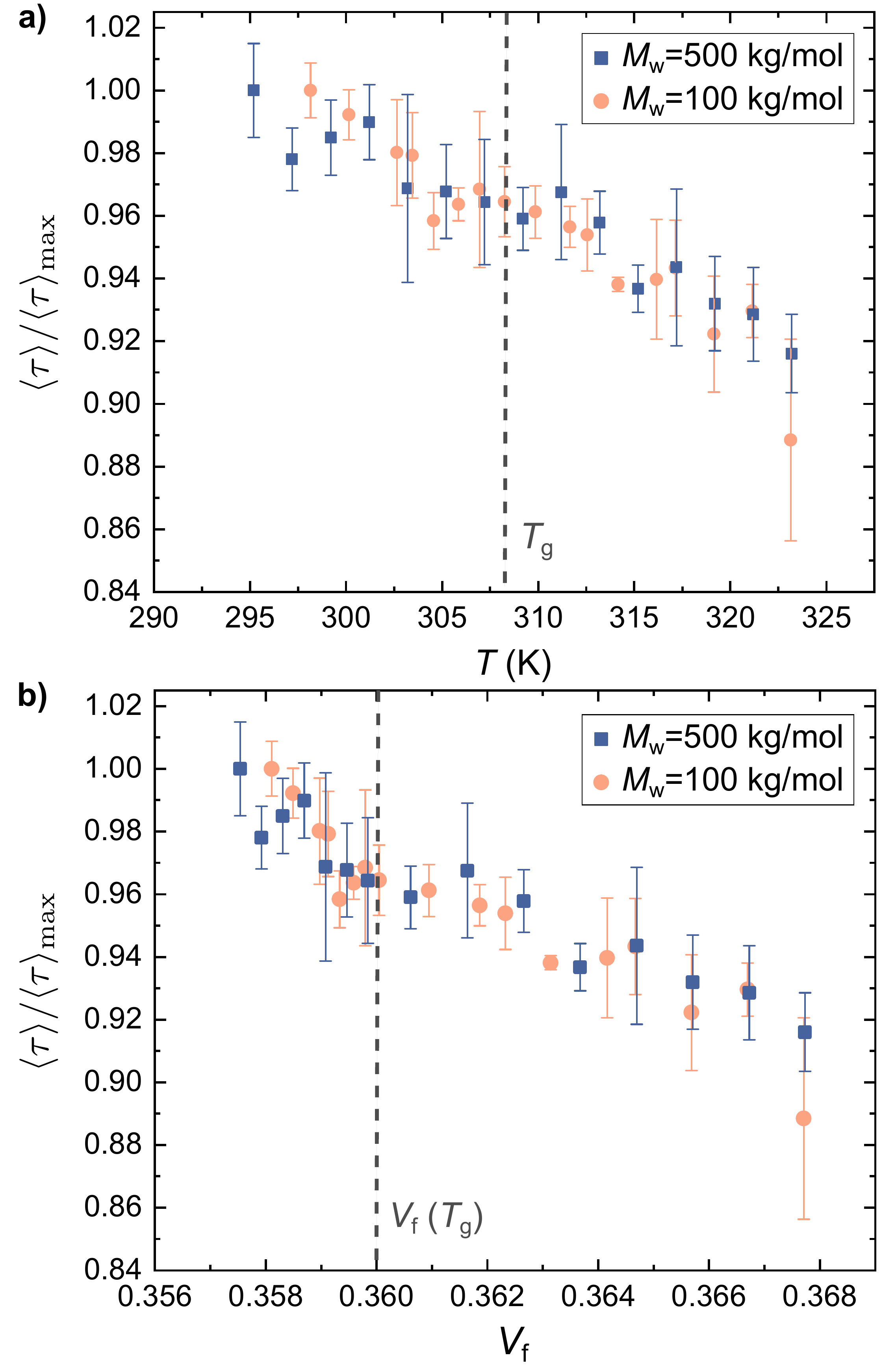} 
    \caption{Normalized lifetime of DCDHF in PVAc polymer glass with two different molecular weight as a function of a) temperature $T$ and b) free volume $V_\mathrm{f}$.}
    \label{fig:PVAc_lifetime}
\end{figure}

\begin{figure*}[ht!]
    \centering
    \includegraphics[width=0.8\textwidth]{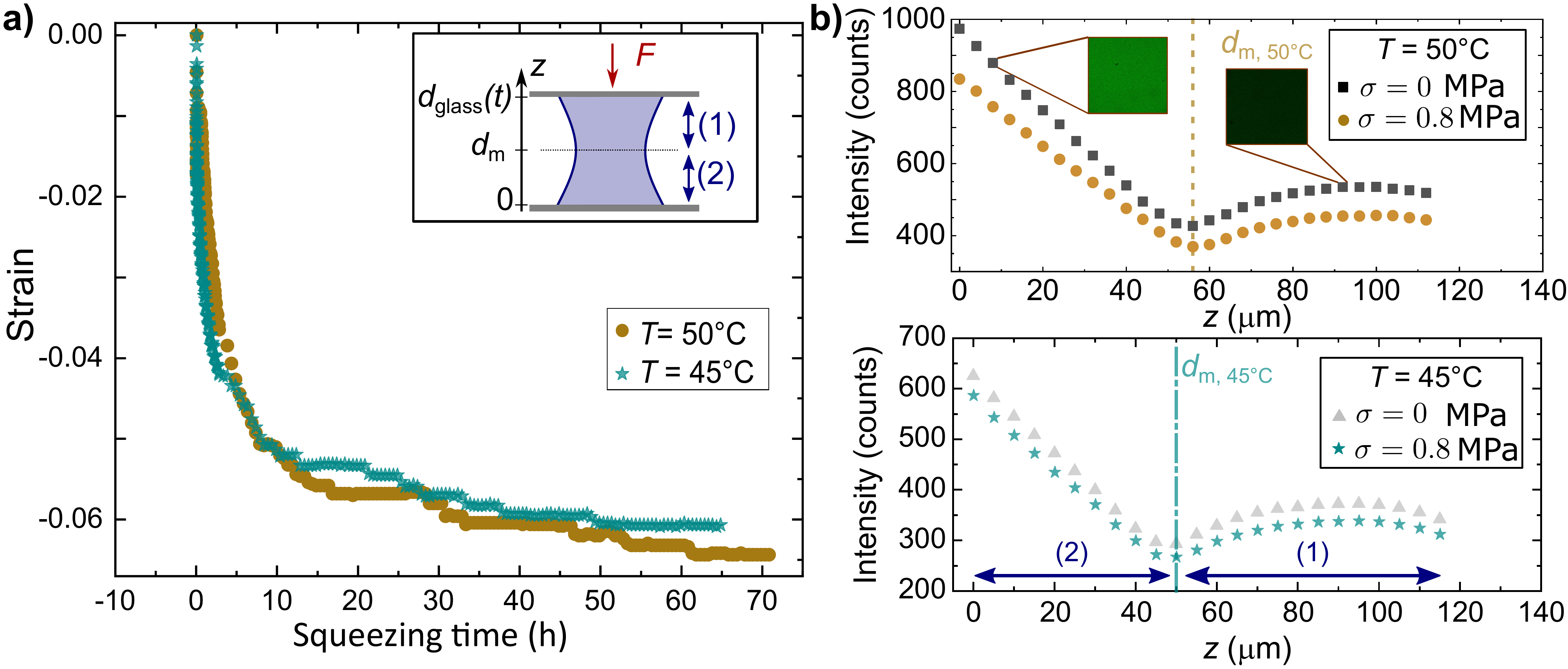} %
    \caption{Squeezing of a polymer \textcolor{black}{film} followed with molecular rotors. a) Deformation of the \textcolor{black}{film} as a function of time under \textcolor{black}{constant stress value of $\sigma = 0.8$\,MPa} . b) Fluorescence intensity as a function of $z$ position in a PVAc polymer \textcolor{black}{film} before and after squeezing. Top: experiment at $T=50^\circ$C ; Bottom: experiment at $T=45^\circ$C . The dotted lines roughly correspond to the position $d_\mathrm{m}$ where the two PVAc glasses (parts (1) and (2) in the inset of Fig.\,\ref{fig:applied_force_comp}a)) were melted together.}%
    \label{fig:applied_force_comp}%
\end{figure*}

Although easier to set up in a laboratory, as it only requires the use of a common confocal microscope, fluorescence intensity measurements are often considered less accurate than fluorescence lifetime measurements. Indeed, the measured intensity may depend on optical parameters, photobleaching, etc. In order to confirm the previous results obtained for the fluorescence intensity, we reproduced the same experiments by measuring the fluorescence lifetime using FLIM microscopy. The additional advantage of lifetime measurement is that it is independent of dye concentration, so that different experiments can be more easily compared. As for the intensity measurements, we observe a decrease of the (amplitude averaged) lifetime with increasing the temperature. As it is depicted in Fig.\,\ref{fig:PVAc_lifetime}a, the normalized lifetime decreases with increasing temperature up to $T = 305\,$K, shows a plateau in the small region around the glass transition temperature and continues to decrease afterwards. The lifetime drops with a slightly smaller slope in comparison to the intensity, possibly due to complex environment surrounding the fluorescence probe \textcolor{black}{or DCDHF may not be in its highest sensitivity range inside PVAc polymer glass}. Fluorescence intensity is rarely systematically compared with fluorescence lifetime for molecular rotors embedded in polymer glass~\cite{anwand_determination_1991}. Such a comparison confirms the potential of \textcolor{black}{DCDHF} to probe the free volume in a \textcolor{black}{PVAc} glass, as simple intensity measurements already provides quantitative information.



\subsection{\textcolor{black}{Shear-banding effect of polymer film under compression}}
We now turn to the measurements of the polymer \textcolor{black}{film} under stress. If a flow is imposed on such a system, either the polymers will have to  rearrange to accommodate the flow, or the system fractures. If a constant stress is imposed on the system, it will either elastically and hence reversibly deform, fracture, or flow, depending on the stress level and time scale of application. In both cases, a flow must lead to a locally enhanced molecular mobility  compared to the unstressed case. The easiest stress to applied is to put a weight onto the top plate confining the \textcolor{black}{film}, hence imposing a constant stress on the material. 


To investigate the effect of \textcolor{black}{compression}, we applied a normal force of $10\,$N (\textcolor{black}{corresponding to a stress $\sigma \sim 0.8$\,MPa}) to PVAc ($M_\mathrm{w} \sim500\,$kg/mol) by rheometer at two different temperatures, $45^\circ$C and $50^\circ$C for more than two days. These relatively high temperatures were necessary to have a  significant flow of the material with this applied force on a reasonable experimental time scale. The inset in Fig.\,\ref{fig:applied_force_comp}a is an illustration of the probed PVAc polymer glass, prepared from two PVAc polymer glasses molten together. The thickness of the polymer \textcolor{black}{film} $d_\mathrm{glass}$ is measured over time with the rheometer and the strain $\left(d_\mathrm{glass}(t)-d_\mathrm{glass}(t=0)\right)/d_\mathrm{glass}(t=0)$ is calculated from these measurements. More quantitatively, this applied normal force causes the sample to flow in the $z$-direction by approximately \textcolor{black}{$6\,\%$ at $45^\circ$C and around $6.4\,\%$ at $50^\circ$C.(Fig.\,\ref{fig:applied_force_comp}a).}   

As we previously observed (section III A) that the fluorescence intensity of DCDHF probes are sensitive to the free volume in the PVAc glass, we follow the changes induced by normal stress under the confocal microscope. \textcolor{black}{Measurements were done typically 10 min. after the end of the squeezing, while the samples were kept at the studied temperature.} As shown in Fig.\,\ref{fig:applied_force_comp}b, for a given applied force and temperature set, the raw intensity values decreases as a function of $z$-position in a non-monotonic behavior. This decrease of intensity must be due to absorption and scattering effect of laser, as well as inhomogeneity of the dye concentration inside the sample, in the $z$ direction. The position $d_\mathrm{m}$ where the two PVAc glasses (parts (1) and (2) in the inset of Fig.\,\ref{fig:applied_force_comp}a) were melted together roughly corresponds to a kink in the intensity data (dotted lines). Both at $T=50^\circ$C (Fig.\,\ref{fig:applied_force_comp}b top) and at $T=45^\circ$C (Fig.\,\ref{fig:applied_force_comp}b bottom), the intensity is always lower after squeezing for each $z$ position scanned, revealing the flow that occurred in the polymer \textcolor{black}{film}.  


\begin{figure}[h]
    \centering
    \includegraphics[width=\columnwidth]{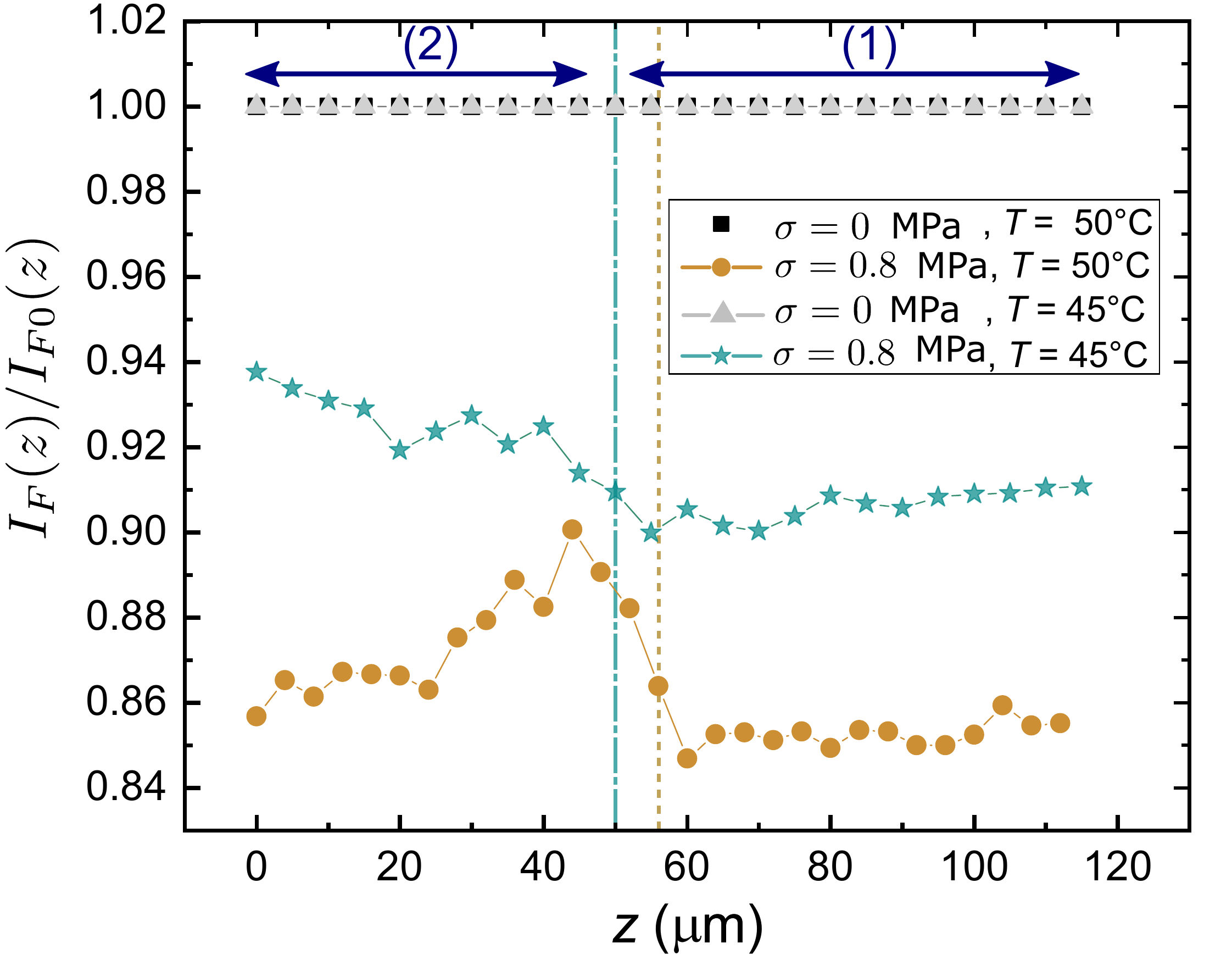} %
    \caption{Normalized intensity $I_{F}(z)/I_{F0}(z)$ as a function of $z$ position in the PVAc  \textcolor{black}{film} after squeezing with \textcolor{black}{constant stress of $\sigma = 0.8$\,MPa} \textcolor{black}{ at $T=45^\circ$C and at $T=50^\circ$C. The fluorescence intensity is measured typically 10 minutes after squeezing the sample.}}%
    \label{fig:applied_force_comp_norm}%
\end{figure}

Fig.\,\ref{fig:applied_force_comp_norm} shows the intensity of the probe after squeezing normalized by the one before squeezing $I_{F}(z)/I_{F0}(z)$ at each $z$ position. The black and grey squares correspond to the normalized intensity $I_{F0}(z)/I_{F0}(z)$ without any squeezing and are by definition at unity. The applied force significantly reduces the normalized intensity with respect to the polymer \textcolor{black}{film} at the same temperature without any external stresses. Interestingly, beside the notable decrease of intensity due to compression of the sample, the normalized intensity value is not homogeneously distributed through the sample in the $z$-direction at either temperature. The normalized fluorescence intensity $I_{F}(z)/I_{F0}(z)$ shows a jump around the position $d_\mathrm{m}$ where the two glasses were molten (dotted lines).  Fig.\,\ref{fig:applied_force_comp_norm} then reveals that the part close to the bottom plate (noted (2) in inset Fig.\,\ref{fig:applied_force_comp}a) has a high fluorescence intensity and hence a high viscosity and low free volume. To the contrary, the part close to the upper plate (noted (1) in inset Fig.\,\ref{fig:applied_force_comp}a), where the stress is applied, shows a very significant decrease of the intensity. The fluorescence probe in the polymer \textcolor{black}{film} therefore reveals an instability that leads to a type of shear banding in polymer \textcolor{black}{film}: one part of the sample has a significantly lower viscosity according to the fluorescence measurements, and hence is likely to flow faster under the constant applied stress. Due to the protocol to prepare thick PVAc glasses, it is not surprising that the shear banding appears close to the position $d_\mathrm{m}$ which is a mechanical weak point. \textcolor{black}{The larger decrease of the normalized intensity at $50^\circ$C must be due to the larger flow of the polymer \textcolor{black}{film} under the external force}. Interestingly, the intensity measurements reveal heterogeneity in the flow in the $z$-direction, namely shear-banding, whereas the ($x,y$) plane stays homogeneous at the microscale (see fluorescence pictures in Fig.\,\ref{fig:applied_force_comp}b)). We therefore clearly observe the direct connection between the macroscopic flow that is detected by the strain measurement with a rheometer and the microscopic detection of this flow by the molecular rotors which detect the microscopic changes of free volume in the polymer film. Unfortunately, due to the extremely long time scales involved it is not possible to perform the similar experiment at glassy state. 

\section{Conclusions}
\textcolor{black}{In this work, we used confocal fluorescence microscopy to measure the fluorescence properties of molecular rotors embedded in a polymer glass. We studied a PVAc polymer glass without any flow near its glass transition temperature, as well as under compression at  $T_\mathrm{g} +10\,$K and  $T_\mathrm{g}+15\,$K. Although fluorescence intensity measurements have been done before, we here do these locally and in addition show that the fluorescence lifetime shows similar behavior as the fluorescence intensity; it confirms that molecular rotors are good dyes to probe changes in free volume at the nanoscale.}  In addition, visualizing the local viscosity in a situation where the \textcolor{black}{polymer film} is forced to flow appears to indicate a shear banding happening in the sample \textcolor{black}{at the microscopic scale. Such heterogeneity in flow at the microscale} will be investigated in more detail \textcolor{black}{in the glassy state} in a future publication.

\begin{acknowledgments}
This project has received funding from the European Research Council (ERC) under the European Union's Horizon 2020 research and innovation program (Grant agreement No. 833240). We thank Felix de Zwart and Bas de Bruin for perfoming DSC measurements.
\end{acknowledgments}

\section*{Data Availability}
The data that support the findings of this study are available from the corresponding author upon reasonable request. 

\bibliographystyle{ieeetr}
\bibliography{ref_PVAc_glass_rev_Marion}

\begin{thebibliography}{10}

\bibitem{kuo_thermal_2003}
S.-W. Kuo, H.-C. Kao, and F.-C. Chang, ``Thermal behavior and specific
  interaction in high glass transition temperature {PMMA} copolymer,'' {\em
  Polymer}, vol.~44, no.~22, pp.~6873--6882, 2003.

\bibitem{alford_specific_1955}
S.~Alford and M.~Dole, ``Specific {Heat} of {Synthetic} {High} {Polymers}.
  {VI}. {A} {Study} of the {Glass} {Transition} in {Polyvinyl} {Chloride},''
  {\em J. Am. Chem. Soc.}, vol.~77, no.~18, pp.~4774--4777, 1955.

\bibitem{rudin_effects_1975}
A.~Rudin and D.~Burgin, ``Effects of molecular weight and chain ends on glass
  transition of polystyrene,'' {\em Polymer}, vol.~16, no.~4, pp.~291--297,
  1975.

\bibitem{pan_conformational_2008}
P.~Pan, B.~Zhu, T.~Dong, K.~Yazawa, T.~Shimizu, M.~Tansho, and Y.~Inoue,
  ``Conformational and microstructural characteristics of poly({L}-lactide)
  during glass transition and physical aging,'' {\em J. Chem. Phys.}, vol.~129,
  no.~18, p.~184902, 2008.

\bibitem{cangialosi_dynamics_2014}
D.~Cangialosi, ``Dynamics and thermodynamics of polymer glasses,'' {\em J.
  Phys.: Condens. Matter}, vol.~26, no.~15, p.~153101, 2014.

\bibitem{vidal_russell_direct_2000}
E.~Vidal~Russell and N.~E. Israeloff, ``Direct observation of molecular
  cooperativity near the glass transition,'' {\em Nature}, vol.~408, no.~6813,
  pp.~695--698, 2000.

\bibitem{balbuena_structural_2021}
C.~Balbuena, M.~M. Gianetti, and E.~R. Soulé, ``A structural study and its
  relation to dynamic heterogeneity in a polymer glass former,'' {\em Soft
  Matter}, vol.~17, no.~12, pp.~3503--3512, 2021.

\bibitem{a_riggleman_heterogeneous_2010}
R.~A.~Riggleman, H.-N. Lee, M.~D.~Ediger, and J.~J.~d. Pablo, ``Heterogeneous
  dynamics during deformation of a polymer glass,'' {\em Soft Matter}, vol.~6,
  no.~2, pp.~287--291, 2010.

\bibitem{yoshimoto_mechanical_2004}
K.~Yoshimoto, T.~S. Jain, K.~V. Workum, P.~F. Nealey, and J.~J. de~Pablo,
  ``Mechanical {Heterogeneities} in {Model} {Polymer} {Glasses} at {Small}
  {Length} {Scales},'' {\em Phys. Rev. Lett.}, vol.~93, no.~17, p.~175501,
  2004.

\bibitem{siekierzycka_polymer_2010}
J.~R. Siekierzycka, C.~Hippius, F.~Würthner, R.~M. Williams, and A.~M.
  Brouwer, ``Polymer {Glass} {Transitions} {Switch} {Electron} {Transfer} in
  {Individual} {Molecules},'' {\em J. Am. Chem. Soc.}, vol.~132, no.~4,
  pp.~1240--1242, 2010.

\bibitem{ellison_confinement_2002}
C.~Ellison, S.~Kim, D.~Hall, and J.~Torkelson, ``Confinement and processing
  effects on glass transition temperature and physical aging in ultrathin
  polymer films: {Novel} fluorescence measurements,'' {\em Eur. Phys. J. E},
  vol.~8, no.~2, pp.~155--166, 2002.

\bibitem{willets_novel_2003}
K.~A. Willets, O.~Ostroverkhova, M.~He, R.~J. Twieg, and W.~E. Moerner, ``Novel
  {Fluorophores} for {Single}-{Molecule} {Imaging},'' {\em J. Am. Chem. Soc.},
  vol.~125, no.~5, pp.~1174--1175, 2003.

\bibitem{lee_dye_2008}
H.-N. Lee, K.~Paeng, S.~F. Swallen, and M.~D. Ediger, ``Dye reorientation as a
  probe of stress-induced mobility in polymer glasses,'' {\em J. Chem. Phys.},
  vol.~128, no.~13, p.~134902, 2008.

\bibitem{lee_molecular_2009}
H.-N. Lee, K.~Paeng, S.~F. Swallen, M.~D. Ediger, R.~A. Stamm, G.~A. Medvedev,
  and J.~M. Caruthers, ``Molecular mobility of poly(methyl methacrylate) glass
  during uniaxial tensile creep deformation,'' {\em Journal of Polymer Science
  Part B: Polymer Physics}, vol.~47, no.~17, pp.~1713--1727, 2009.

\bibitem{lee_direct_2009}
H.-N. Lee, K.~Paeng, S.~F. Swallen, and M.~D. Ediger, ``Direct {Measurement} of
  {Molecular} {Mobility} in {Actively} {Deformed} {Polymer} {Glasses},'' {\em
  Science}, vol.~323, no.~5911, pp.~231--234, 2009.

\bibitem{lee_interaction_2010}
H.-N. Lee and M.~D. Ediger, ``Interaction between physical aging, deformation,
  and segmental mobility in poly(methyl methacrylate) glasses,'' {\em J. Chem.
  Phys.}, vol.~133, no.~1, p.~014901, 2010.

\bibitem{hebert_reversing_2017}
K.~Hebert and M.~D. Ediger, ``Reversing {Strain} {Deformation} {Probes}
  {Mechanisms} for {Enhanced} {Segmental} {Mobility} of {Polymer} {Glasses},''
  {\em Macromolecules}, vol.~50, no.~3, pp.~1016--1026, 2017.

\bibitem{bending_comparison_2016}
B.~Bending and M.~D. Ediger, ``Comparison of mechanical and molecular measures
  of mobility during constant strain rate deformation of a {PMMA} glass,'' {\em
  Journal of Polymer Science Part B: Polymer Physics}, vol.~54, no.~19,
  pp.~1957--1967, 2016.

\bibitem{lee_deformation-induced_2009}
H.-N. Lee, R.~A. Riggleman, J.~J. de~Pablo, and M.~D. Ediger,
  ``Deformation-{Induced} {Mobility} in {Polymer} {Glasses} during {Multistep}
  {Creep} {Experiments} and {Simulations},'' {\em Macromolecules}, vol.~42,
  no.~12, pp.~4328--4336, 2009.

\bibitem{riggleman_free_2007}
R.~A. Riggleman, H.-N. Lee, M.~D. Ediger, and J.~J. de~Pablo, ``Free {Volume}
  and {Finite}-{Size} {Effects} in a {Polymer} {Glass} under {Stress},'' {\em
  Phys. Rev. Lett.}, vol.~99, no.~21, p.~215501, 2007.

\bibitem{hall_translation-rotation_1997}
D.~B. Hall, A.~Dhinojwala, and J.~M. Torkelson, ``Translation-{Rotation}
  {Paradox} for {Diffusion} in {Glass}-{Forming} {Polymers}: {The} {Role} of
  the {Temperature} {Dependence} of the {Relaxation} {Time} {Distribution},''
  {\em Phys. Rev. Lett.}, vol.~79, no.~1, pp.~103--106, 1997.

\bibitem{lee_mechanical_2010}
H.-N. Lee and M.~D. Ediger, ``Mechanical {Rejuvenation} in {Poly}(methyl
  methacrylate) {Glasses}? {Molecular} {Mobility} after {Deformation},'' {\em
  Macromolecules}, vol.~43, no.~13, pp.~5863--5873, 2010.

\bibitem{salmon_velocity_2003}
J.-B. Salmon, A.~Colin, S.~Manneville, and F.~Molino, ``Velocity {Profiles} in
  {Shear}-{Banding} {Wormlike} {Micelles},'' {\em Phys. Rev. Lett.}, vol.~90,
  no.~22, p.~228303, 2003.

\bibitem{radulescu_time_2003}
O.~Radulescu, P.~D. Olmsted, J.~P. Decruppe, S.~Lerouge, J.-F. Berret, and
  G.~Porte, ``Time scales in shear banding of wormlike micelles,'' {\em
  Europhysics Letters}, vol.~62, no.~2, p.~230, 2003.

\bibitem{varnik_study_2004}
F.~Varnik, L.~Bocquet, and J.-L. Barrat, ``A study of the static yield stress
  in a binary {Lennard}-{Jones} glass,'' {\em J. Chem. Phys.}, vol.~120, no.~6,
  pp.~2788--2801, 2004.

\bibitem{tapadia_banding_2006}
P.~Tapadia, S.~Ravindranath, and S.-Q. Wang, ``Banding in {Entangled} {Polymer}
  {Fluids} under {Oscillatory} {Shearing},'' {\em Phys. Rev. Lett.}, vol.~96,
  no.~19, p.~196001, 2006.

\bibitem{cao_shear_2012}
J.~Cao and A.~E. Likhtman, ``Shear {Banding} in {Molecular} {Dynamics} of
  {Polymer} {Melts},'' {\em Phys. Rev. Lett.}, vol.~108, no.~2, p.~028302,
  2012.

\bibitem{wisitsorasak_dynamical_2017}
A.~Wisitsorasak and P.~G. Wolynes, ``Dynamical theory of shear bands in
  structural glasses,'' {\em PNAS}, vol.~114, no.~6, pp.~1287--1292, 2017.

\bibitem{ravindranath_banding_2008}
S.~Ravindranath, S.-Q. Wang, M.~Olechnowicz, and R.~P. Quirk, ``Banding in
  {Simple} {Steady} {Shear} of {Entangled} {Polymer} {Solutions},'' {\em
  Macromolecules}, vol.~41, no.~7, pp.~2663--2670, 2008.

\bibitem{loutfy_effect_1983}
R.~O. Loutfy and D.~M. Teegarden, ``Effect of polymer chain tacticity on the
  fluorescence of molecular rotors,'' {\em Macromolecules}, vol.~16, no.~3,
  pp.~452--456, 1983.

\bibitem{law_fluorescence_1983}
K.~Y. Law and R.~O. Loutfy, ``Fluorescence probe for microenvironments: {On}
  the fluorescence properties of p-{N},{N}-dialkylaminobenzylidenemalononitrile
  in polymer matrices,'' {\em Polymer}, vol.~24, no.~4, pp.~439--442, 1983.

\bibitem{loutfy_high-conversion_1981}
R.~O. Loutfy, ``High-conversion polymerization of fluorescence probes. 1.
  {Polymerization} of methyl methacrylate,'' {\em Macromolecules}, vol.~14,
  no.~2, pp.~270--275, 1981.

\bibitem{priestley_effects_2007}
R.~D. Priestley, M.~K. Mundra, N.~J. Barnett, L.~J. Broadbelt, and J.~M.
  Torkelson, ``Effects of {Nanoscale} {Confinement} and {Interfaces} on the
  {Glass} {Transition} {Temperatures} of a {Series} of {Poly}(n-methacrylate)
  {Films},'' {\em Australian Journal of Chemistry}, vol.~60, no.~10,
  pp.~765--771, 2007.

\bibitem{kim_effect_2008}
S.~Kim, C.~B. Roth, and J.~M. Torkelson, ``Effect of nanoscale confinement on
  the glass transition temperature of free-standing polymer films: {Novel},
  self-referencing fluorescence method,'' {\em Journal of Polymer Science Part
  B: Polymer Physics}, vol.~46, no.~24, pp.~2754--2764, 2008.

\bibitem{kim_confinement_2009}
S.~Kim, S.~A. Hewlett, C.~B. Roth, and J.~M. Torkelson, ``Confinement effects
  on glass transition temperature, transition breadth, and expansivity:
  {Comparison} of ellipsometry and fluorescence measurements on polystyrene
  films,'' {\em Eur. Phys. J. E}, vol.~30, no.~1, p.~83, 2009.

\bibitem{loutfy_effect_1982}
R.~O. Loutfy and B.~A. Arnold, ``Effect of viscosity and temperature on
  torsional relaxation of molecular rotors,'' {\em J. Phys. Chem.}, vol.~86,
  no.~21, pp.~4205--4211, 1982.

\bibitem{law_spectroscopy_1981}
K.~Y. Law and R.~O. Loutfy, ``Spectroscopy of dyes in polymer matrixes: dual
  fluorescence of p-[(dialkylamino)benzylidene]malononitrile,'' {\em
  Macromolecules}, vol.~14, no.~3, pp.~587--591, 1981.

\bibitem{hooker_coupling_1995}
J.~C. Hooker and J.~M. Torkelson, ``Coupling of {Probe} {Reorientation}
  {Dynamics} and {Rotor} {Motions} to {Polymer} {Relaxation} {As} {Sensed} by
  {Second} {Harmonic} {Generation} and {Fluorescence},'' {\em Macromolecules},
  vol.~28, no.~23, pp.~7683--7692, 1995.

\bibitem{royal_monitoring_1990}
J.~S. Royal and J.~M. Torkelson, ``Monitoring the molecular scale effects of
  physical aging in polymer glasses with fluorescence probes,'' {\em
  Macromolecules}, vol.~23, no.~14, pp.~3536--3538, 1990.

\bibitem{royal_molecular-scale_1992}
J.~S. Royal and J.~M. Torkelson, ``Molecular-scale asymmetry and memory
  behavior in poly(vinyl acetate) monitored with mobility-sensitive fluorescent
  molecules,'' {\em Macromolecules}, vol.~25, no.~6, pp.~1705--1710, 1992.

\bibitem{royal_physical_1993}
J.~S. Royal and J.~M. Torkelson, ``Physical aging effects on molecular-scale
  polymer relaxations monitored with mobility-sensitive fluorescent
  molecules,'' {\em Macromolecules}, vol.~26, no.~20, pp.~5331--5335, 1993.

\bibitem{suhina_photophysics_2021}
T.~Suhina, D.~Bonn, B.~Weber, and A.~M. Brouwer, ``Photophysics of
  {Fluorescent} {Contact} {Sensors} {Based} on the {Dicyanodihydrofuran}
  {Motif},'' {\em Chemphyschem}, vol.~22, no.~2, pp.~221--227, 2021.

\bibitem{weber_molecular_2018}
B.~Weber, T.~Suhina, T.~Junge, L.~Pastewka, A.~M. Brouwer, and D.~Bonn,
  ``Molecular probes reveal deviations from {Amontons}’ law in multi-asperity
  frictional contacts,'' {\em Nat Commun}, vol.~9, no.~1, p.~888, 2018.

\bibitem{lord_dcdhf_2009}
S.~J. Lord, N.~R. Conley, H.-l.~D. Lee, S.~Y. Nishimura, A.~K. Pomerantz, K.~A.
  Willets, Z.~Lu, H.~Wang, N.~Liu, R.~Samuel, R.~Weber, A.~Semyonov, M.~He,
  R.~J. Twieg, and W.~E. Moerner, ``{DCDHF} {Fluorophores} for
  {Single}-{Molecule} {Imaging} in {Cells},'' {\em Chemphyschem}, vol.~10,
  no.~1, pp.~55--65, 2009.

\bibitem{jee_internal_2010}
A.-Y. Jee, E.~Bae, and M.~Lee, ``Internal motion of an electronically excited
  molecule in viscoelastic media,'' {\em J. Chem. Phys.}, vol.~133, no.~1,
  p.~014507, 2010.

\bibitem{jee_determination_2013}
A.-Y. Jee, H.~Lee, Y.~Lee, and M.~Lee, ``Determination of the elastic modulus
  of poly(ethylene oxide) using a photoisomerizing dye,'' {\em Chemical
  Physics}, vol.~422, pp.~246--250, 2013.

\bibitem{saito_temperature_1963}
S.~Saito, ``Temperature dependence of dielectric relaxation behavior for
  various polymer systems,'' {\em Kolloid-Z.u.Z.Polymere}, vol.~189, no.~2,
  pp.~116--125, 1963.

\bibitem{zaccone_disorder-assisted_2013}
A.~Zaccone and E.~M. Terentjev, ``Disorder-{Assisted} {Melting} and the {Glass}
  {Transition} in {Amorphous} {Solids},'' {\em Phys. Rev. Lett.}, vol.~110,
  no.~17, p.~178002, 2013.

\bibitem{bogoslovov_effect_2008}
R.~B. Bogoslovov, C.~M. Roland, A.~R. Ellis, A.~M. Randall, and C.~G.
  Robertson, ``Effect of {Silica} {Nanoparticles} on the {Local} {Segmental}
  {Dynamics} in {Poly}(vinyl acetate),'' {\em Macromolecules}, vol.~41, no.~4,
  pp.~1289--1296, 2008.

\bibitem{anwand_determination_1991}
D.~Anwand, F.~W. Müller, B.~Strehmel, and K.~Schiller, ``Determination of the
  molecular mobility and the free volume of thin polymeric films with
  fluorescence probes,'' {\em Die Makromolekulare Chemie}, vol.~192, no.~9,
  pp.~1981--1991, 1991.

\end{thebibliography}
\end{document}